\documentclass[english,twocolumns]{IEEEtran}
\usepackage[T1]{fontenc}
\usepackage{amsmath}
\usepackage{setspace}
\usepackage{amssymb}
\usepackage{multicol}
\usepackage{graphicx}
\usepackage{cite}
\usepackage{xcolor}

\makeatletter
\newtheorem{df}{Definition}
\newtheorem{thm}{Theorem}
\newtheorem{prop}{Proposition}
\newtheorem{lemma}{Lemma}
\newtheorem{example}{Example}

\newtheorem{cor}{Corollary}
\newtheorem{remark}{Remark}
\makeatother

\usepackage{babel}

\newcommand{\red}[1]{{\color{red} #1 \color{black}}}

\begin{document}

\def\N{{\cal N}}
\def\bz{{z_{\scriptscriptstyle \N}}}
\def\by{{y_{\scriptscriptstyle \N}}}
\def\l{\left}
\def\r{\right}
\def\gf{ {\mathbb F}}
\def\rankfn{\rho}
\def\A{\mathcal A}
\def\M{\mathcal M}
\def\X{\mathcal X}
\def\B{\mathcal B}
\def\reals{\mathbb R}
\def\Z{\mathcal Z}
\def\K{\mathcal K}
\def\Q{\mathcal Q}
\def\C{\mathcal C}
\def\p{\prime}
\def\real{{\mathbb R}}
\def\dist{{W}}

\def\Cd{{\C^{\perp}}}
\def\indicator{{P}}
\def\T{{P^{[2]}}}
\def\sigmaa{{S}}
\def\bfz{{\bf z}}
\def\perm{{\Omega}}

\def\decode{{g}}
\def\pir{{\Lambda}}
\def \storagecost{{\mathsf {SC}}}
\def\downloadcost{{\mathsf{RC}}}

\def\xsize{{r}}
\def\dsize{{s}}

\def\mC{{{C}}}
\def\mD{{{\bf D}}}
\def\mX{{{\bf X}}}
\def\event{{\mathbb E}}
 
\def\e{{\epsilon}}

\newcounter{tempEquationCounter}
\newcounter{thisEquationNumber}
\newenvironment{floatEq}
{\setcounter{thisEquationNumber}{\value{equation}}\addtocounter{equation}{1}
\begin{figure*}[!t]
\normalsize\setcounter{tempEquationCounter}{\value{equation}}
\setcounter{equation}{\value{thisEquationNumber}}
}
{\setcounter{equation}{\value{tempEquationCounter}}
\hrulefill\vspace*{4pt}
\end{figure*}

}

%

\title{Private Information Retrieval for Coded  Storage}

\author{
Terence H. Chan, Siu-Wai Ho, and Hirosuke Yamamoto

\IEEEauthorblockN{
  }

}

%



\maketitle
\newcommand{\nle}[1]{\stackrel{#1}{\le}}
\newcommand{\nge}[1]{\stackrel{#1}{\ge}}
\newcommand{\nequal}[1]{\stackrel{#1}{=}}
\def\O{{ c' (c-1/2) H(D_{n^{*}})   }}

\def\calJ{{\cal J}}
\def\support{{\Lambda}}
\def\br{{\bf r}}
\def\bc{{\bf c}}
\def\bz{{\bf z}}
\def\bs{{\bf s}}
\def\bu{{\bf u}}
\def\bv{{\bf v}}
\def\bg{{\bf g}}
\def\bff{{\bf f}}
\def\ground{{\Omega}}
\newcommand\base[1]{{{E}_{#1}}}
\def\tN{{M}}

\def\S{{\cal S}}
\def\L{{\cal L}}
\def\R{{\cal R}}
\def\T{{\cal T}}

\def\bc{{\bf c}}

\def\solA{{\bf A}}
\def\solB{{\bf B}}
\def\solC{{\bf C}}
\def\field{{\mathbb F}_{q}}
\def\Kp{{\mathcal K}^{'}}

\def\J{{\mathcal J}}
\def\bfy{{\bf y}}


\def\mM{{\bf M}}
\def\mP{{\bf P}}
\def\mU{{\bf U}}
\def\mC{{\bf X}}
\def\mV{{\bf V}}
\def\mW{{\bf W}}

\def\vu{{\bf u}}
\def\vx{{\bf x}}
\def\vd{{\bf d}}
\def\vv{{\bf v}}

\def\vs{{\bf s}}

\def\vz{{\bf z}}

\def\storage{{\Omega}}

\newcommand\rank[1]{{{\bf rank}(#1)}}

\begin{abstract}
Private information retrieval scheme for coded data storage is considered in this paper.
We focus on the case where the size of each data record is large and hence only the download cost (but not the upload cost for transmitting retrieval queries) is of interest. 
We prove that the tradeoff between storage cost and retrieval/download cost depends on the number of data records in the system.
We also propose a fairly general class of linear storage codes and retrieval schemes and derive conditions under which our retrieval schemes are error-free and private. Tradeoffs between the storage cost and retrieval costs are also obtained. 
Finally,  we consider special cases when the underlying storage code is based on an MDS code. Using our proposed method, we show that a randomly generated retrieval scheme is indeed very likely to be private and error-free.  
\end{abstract}


\section{Introduction}
%
%
%
%
%
%
%
%
%
%
%
%
%
%
%

In modern data storage system, data are usually stored at multiple storage nodes in the cloud. However, system failures are inevitable and in fact common in modern information-technological infrastructure. 
These failures can be caused by hardware or software failures, or even natural disaster (such as fires, earthquakes and flooding).
Another big challenge is how to ensure data stored in the cloud are protected from loss and from being overwritten by illegitimate parties.
To ensure that data will not be lost in the case of failures, data redundancy are required.
Traditionally, several  copies of the same  data will be stored. Thus, if one of the copies stored in a node is lost, the lost content  can be recovered by retrieving a copy of the data from other surviving nodes.

Such  replication approach is not efficient in terms of storage cost. An alternative approach is to use erasure codes (such as Reed-Solomon codes, an example of maximum distance separable codes) to store the data. Thus, any content loss  can then be seen as a data erasure  and  be recovered from the code.
Comparing with replication,  this coding approach has a lower storage cost but also a higher repair cost (measured by the amount of data sent to repair a failure node).
There is a fundamental tradeoff between the costs for storage and repair. In \cite{Dimakis2010Network}, regenerating codes were proposed  to achieve the optimal tradeoff between storage cost and repair cost. This approach has since received a lot of attentions ~\cite{Thakor2013Symmetry,Tebbi2014Linear,Rashmi_Shah_Kumar_Ramchandran_2009,Shah_Rashmi_Kumar_Ramchandran_2009}.

In some data storage applications, not only the data protection against system failure and illegitimate parties is a concern but also the privacy of data retrieval query needs to be protected.
 For example, consider investors retrieving stock prices from a cloud storage provider (CSP). In this case, it is preferred to keep  the identity of the stocks secret from  other users  and  the CSP.
Private information retrieval (PIR) ensures that servers storing data would not know the identity of the  data record that a user is retrieving.
One naive approach to achieve PIR is by downloading every record in the system no matter which record the user is needed.   The downside of this approach is the extremely large retrieval cost, which linearly increases with $N$ (the number of records stored).

The challenge however is how to design an efficient PIR scheme (in terms of costs for data storage and retrieval).
In a $k$-server PIR scheme, a user sends $k$ queries to $k$ servers which will reply to the user $k$ answers. 
The retrieval efficiency of a PIR scheme is measured by  the upload cost (i.e., the size of the queries) and the retrieval cost (i.e., the size of the answers). 
The sum of these two costs is referred to as the total communication complexity.
In \cite{Chor1995}, two very interesting schemes were proposed.
The first scheme achieves total communication complexity with $O(N^{\frac{1}{K}})$ where $K$  is the number of servers and $N$ is the number of data records.
If there are $\frac{1}{3} \log_2(N) + 1$ servers,
the second scheme achieves a total communication complexity with
$\frac{1}{3}(1 + o(1)) \cdot \log_2^2(N) \cdot log_2(log_2(2N))$.
These results were further elaborated in  \cite[Sec. 3.1-3.4]{Chor1998}.
For data records of size equal to $\ell$, a PIR scheme with total communication complexity equal to $4 \ell$ was proposed.
Note that all these PIR schemes require that each storage node stores a copy of every data records. 
This paper considers a more general scenario where this requirement is not necessarily satisfied.

In \cite{Kushilevitz1997Replication}, a single-database ``computational'' PIR scheme to achieve retrieval cost less than $N$ was proposed. Unlike previous information-theoretic schemes, the privacy of the identity of the retrieved data is protected based on the computational hardness of the quadratic residuosity problem. This scheme can achieve $N^\epsilon$ for any $\epsilon > 0$ by assuming the hardness of deciding quadratic residuosity.
In  \cite{Yekhanin2007Locally,Gentry815Single-Database}, PIR schemes which   achieve a retrieval cost of $\log^2 N$ were proposed.
However, the computational complexity of these schemes is of  concern.
In fact, a scheme which requires less computation of public-key operations was shown in \cite{Lipmaa2009First}.
Information-theoretic approach has also been considered to achieve PIR \cite{Beimel2002Robust}.
A more general model was considered in \cite{Devet2012Optimally} in which  1) not every server may respond to the query and 2) some servers are tampered and hence erroneous responses will be returned to the users.
There are many other works which considered similar PIR problems. Due to  limited space,  please refer to \cite{Chor1998} for more details.

Almost all of the existing works on PIR assume uncoded data storage (where each storage node stores a copy of every data record). However, uncoded storage (or more precisely, storage based on repetition codes) has the highest storage cost. To reduce the storage cost, coded storage (e.g., ones that base on maximum distance separable (MDS) codes, or locally repairable codes) have been proposed. Motivated by the advances in coded data storage, this paper considers PIR schemes for coded data storage system.

To the best of our knowledge, the recent work\cite{Shah2014One-Extra} is the only existing work that discusses private information retrieval in coded storage. However, there are some fundamental differences between the approach in \cite{Shah2014One-Extra} and the one considered in this paper. In \cite{Shah2014One-Extra}, it was proved that only one extra bit is required to download to ensure the privacy of the identity of retrieved record. However, the extremely low retrieval cost comes with a price that the number of storage nodes (and also the storage cost) needs to grow with the data record size. We however assume that the number of storage nodes is fixed. 

In \cite[Section V]{Shah2014One-Extra}, PIR schemes for MDS codes based storage were proposed. The optimality of these schemes were not discussed. In contrast, this paper considers a fairly general framework where  we impose no restriction on the choice of the underlying storage codes. We obtain the tradeoffs between storage cost and retrieval cost and show that the optimal tradeoff can be achieved if MDS codes are used. Unlike in \cite{Shah2014One-Extra} where the number of data records is assumed to be fixed,  we will also show that the optimal tradeoff between the storage and retrieval costs also depend on the number of records in the system.


The organisation of the paper is as follows.
Section \ref{sec:2} is   the background for storage codes and private information retrieval (PIR) scheme. In Section \ref{sec:3}, we  propose our PIR schemes for coded storage. Sufficient conditions for which our scheme is private  will be derived. In Section \ref{sec:tradeoff},  we will identify the tradeoff between storage cost and retrieval cost.

\subsubsection*{Notations}
For random variables $X,Y$, the entropy of  $X$ will be denoted by $H(X)$, while the mutual information between $X$ and $Y$ is denoted by $I(X;Y)$. A set of random variables $(X_{i}, i \in \alpha) $ can also be denoted by $X_{\alpha}$. Random variables are usually denoted by capital letters (such as $X,Y$ etc) while  their realisations by small letters. Sample spaces over which random variables are defined will be denoted by calligraphic letters.

\section{Problem Formulation}\label{sec:2}
\def\enc{{\Omega}}
\newcommand\I[2]{{ {I} ( #1 ; #2 )  }}

\subsection{System Formulation}
\def\rp{{n^{*}}} 

We assume that  there are $N$ data records $(D_{n}, \: n \in \N \triangleq \{1, \ldots,  {N}\})$ to be stored in 
$K$ data storage nodes  (indexed by   ${\cal K} \triangleq \{1, \ldots, K\}$). Assume without loss of generality that  these data records are  independently and uniformly distributed over a  common sample space   ${\cal D}$. In other words, 
\begin{align}
\Pr(D_{n}=d_{n}, n \in \N) & =  1/|{\cal D}|^{N}. 
\end{align}
Hence, 
$
H(D_{\N}) = \sum_{n\in\N}H(D_{n})
$
and
$
H(D_{n}) = \log |{\cal D}|
$.

\begin{df}[Storage Code]
For any $k\in\K$, let $X_{k}$ (defined over ${\cal X}_{K}$) be the content stored in storage node $k$. 
A  $(N,K )$  \emph{storage code}   is defined  by an encoding mapping
\[
\enc : \: {\cal D}^{N} \rightarrow \prod_{k\in{\cal K}}{\cal X}_{k}
\]
such that 
$
(X_{1}, \ldots, X_{K}) = \enc(D_{\N})
$.
\end{df}

Clearly, it is required that all data $D_{1}, \ldots, D_{N}$ can be retrieved from the contents in the $K$ storage nodes. In other words, $H(D_{1}, \ldots, D_{N} | X_1, \ldots, X_{K})    =  0$.
In addition,  there are often other criteria that a storage code needs to  satisfy. 
For instance, contents stored at a storage node  may be corrupted (due to software or hardware failures) and hence need to be  recovered from  other  storage nodes. 
For example, one may require that  for any $k\in \K$, there exists a subset $\alpha \subseteq \K\setminus \{k\}$ of size at most $r$ such that 
$
H(X_{k} | X_{j} , j \in \alpha) = 0
$~\cite{Tebbi2014Linear,papailiopoulos2012locally}.
In other scenarios,  it may also be required that all data can still be recovered when there are multiple (no more than $\delta$) simultaneous node failures.  In this case, it is required that 
\[
H(X_{\K} | X_{j}, j \in \beta) = 0
\]
for any $\beta \subseteq \K$ of size $K - \delta$.

For any storage code $\storage$, we will use $\storagecost(\storage)$ to denote the 
 \emph{storage cost}, which is  the number of  bits each storage node must use to store one information bit. More precisely,  
\begin{align}\label{eq2}
\storagecost(\storage) \triangleq   \max_{k\in\K} \frac{  \log |{\cal X}_{k} |}{ N \log |{\cal D}| }.
\end{align}

\begin{remark}
We do not assume any compression schemes used in the storage and hence  $ \log |{\cal X}_{k} |$ is used instead of $H(X_{k})$. As $ \log |{\cal X}_{k} | \ge H(X_{k})$, results obtained in this paper will still hold if $H(X_{k})$ is used. 
Also, there are other alternative storage cost measures (e.g.,  ``the total storage cost'' ${ \sum_{k\in\K}  \log |{\cal X}_{k} |}/{ N \log |{\cal D}| }$). We consider \eqref{eq2} to focus on the required storage cost on individual nodes. With respect to our code (to be defined in Section  \ref{sec:3}), the two measures differed only by a constant multiplicative  factor  $K$ (which is the number of storage nodes and is fixed).
\end{remark}

\begin{figure}
  \begin{center}
        \includegraphics[scale=0.5]{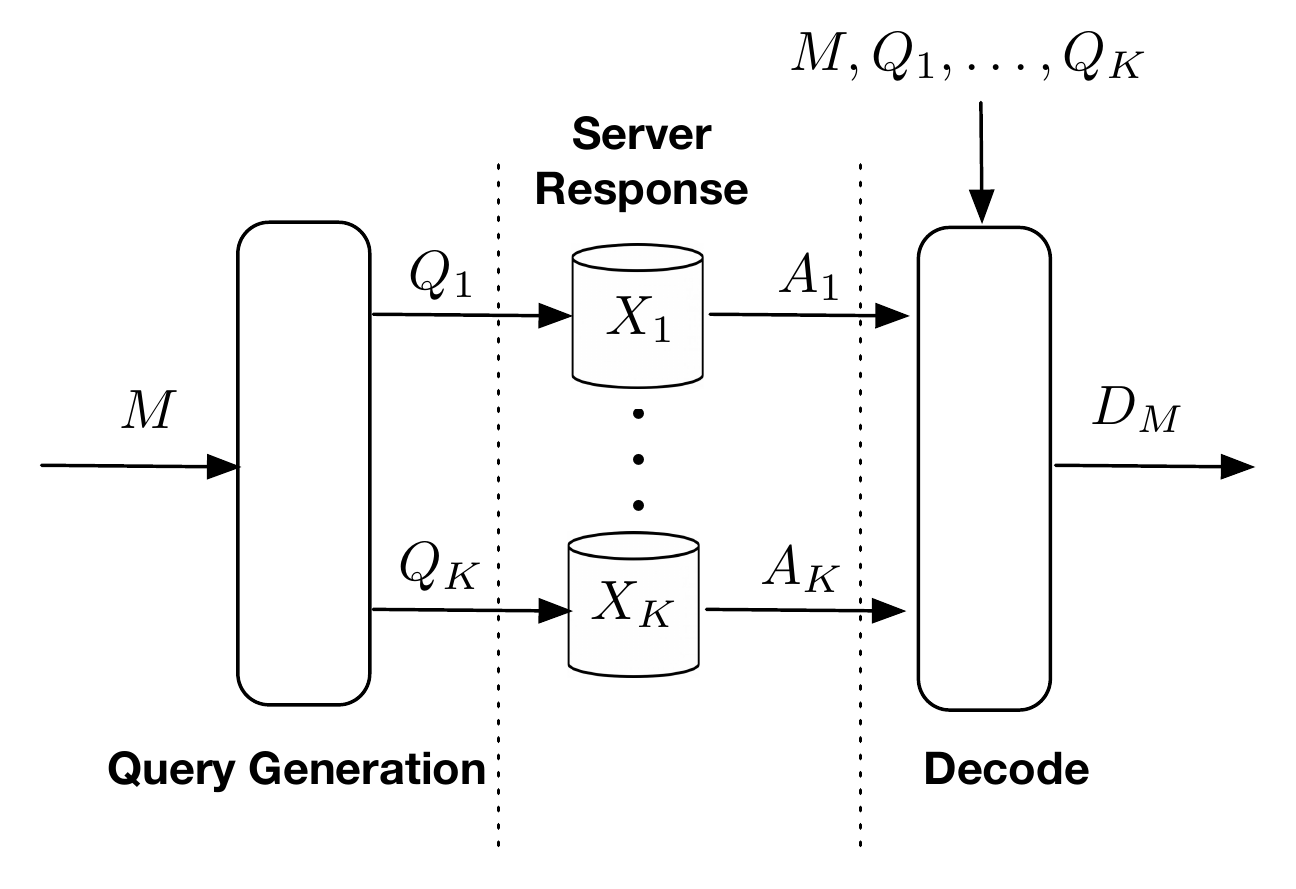} \caption{\label{fig:PIRChart}
 Retrieving data record $D_{M}$}
  \end{center}
\end{figure}

\def\tamper{{\Phi}}

Clearly, a  data storage will be useless if the stored data cannot be retrieved.  
A \emph{retrieval scheme}   usually consists of three steps (see Figure \ref{fig:PIRChart}). The first step is the \emph{queries generation}.
Let $\tN \in \N$ be the index of the data record to be retrieved. We will assume that $\tN$ is uniformly distributed over $\N$.
To retrieve $D_{\tN}$, a user will first generate $K$ queries
$(Q_{k}, k \in\K )$ according to a conditional probability distribution\footnote{
In this paper, we consider the general case that queries are randomly generated to prevent the storage nodes from  knowing the identity of the data record to be retrieved.
}
\begin{align}\label{eq5}
\Lambda_{1}(q_{k}, k\in \K | m) =\Pr(Q_{k} = q_{k}, k\in \K | M=m).  
\end{align}
Then, the query $Q_{k}$  will  be sent  to storage node $k$.
We call $\Lambda_{1}$ the ``\emph{query generation distribution}''.

In the second step, each storage node $k$, receiving $Q_{k}$, will generate a response via the function
\begin{align}
\Lambda_{2,k}: \: \X_{k} \times \Q_{k} \rightarrow \A_{k}. \label{eq:genA}
%
%
\end{align}
In particular, 
$A_{k} \triangleq \Lambda_{2,k} (X_{k}, Q_{k}) $
will be computed and returned to the user. The mappings
$(\Lambda_{2,k}, k\in \K)$  are called ``\emph{response mappings}''.
Finally, in the third step, the user will ``\emph{decode}'' $D_{\tN}$ from
$(\tN, Q_{k}, A_{k}, k\in \K)$.

\begin{df}A retrieval scheme $\Lambda$ is specified by  the query generation distribution and the response mappings 
$$  \Lambda\triangleq (\Lambda_{1} , \Lambda_{2,k}, k\in \K).
$$
\end{df}

\begin{remark} Once the storage code $\storage$ and the retrieval scheme $\Lambda$ are given, the probability distribution of $(M,X_{k}, Q_{k}, A_{k}, k\in \K, D_{n}, n\in\N)$ is well-defined. In particular,
%
%
\begin{align}
(X_{\cal K}, D_{\cal N}) \perp (\tN,  Q_{\cal K})  \label{remark2.a}\\
H(A_{k} |  Q_{k}, X_{k})  = 0 \label{remark2.c} \\
H(D_{n}, n\in \N ) = \sum_{n\in\N} H(D_{n}) .\label{remark2.d}
\end{align}

Conversely, if the probability distribution of 
$(X_{k}, k\in \K, D_{n}, n\in\N)$ is known, then one can also deduce the underlying storage code $\storage$. Similarly, the probability distributions of $(M, Q_{k}, k\in \K)$  (and $(Q_{k},A_{k}, X_{k})$) respectively specify $\Lambda_{1}$ (and $\Lambda_{2,k}$). Therefore, in this paper, we often refer a storage system  by the set of random variables  $(M,X_{k}, Q_{k}, A_{k}, k\in \K, D_{n}, n\in\N)$.

%
\end{remark}
%
%
%
%
%
%
%

\begin{df}[Error-free Retrieval]
For a given storage code $\storage$, a retrieval scheme $\Lambda$  is called \emph{error-free} if 
\begin{align}\label{def3.eq}
H(D_{\tN} | \tN, Q_{\K} , A_{\K} ) = 0.
\end{align}
\end{df}

\begin{df}[Private Retrieval]
Let $\tamper$ be a collection of subsets of $\K$. A retrieval scheme $\Lambda$  is called $\tamper$-\emph{private} if 
\begin{align}
I(\tN ; Q_{j}, j\in \alpha)  = 0 \label{def4:eq.a}
\end{align}
for all subset $\alpha \in \tamper$. 
We call  $\alpha$ the \emph{collusion pattern}. 
\end{df}

According to our definitions, if a retrieval scheme is error-free and $\tamper$-\emph{private}, then 1) a user can always retrieve (and decode) its interested data record $D_{\tN}$ with no errors, and 2)  for any $\alpha \in \tamper$, the colluding set of storage nodes (i.e., indexed by $\alpha$) obtains no information about the identity (the index $M$) but not necessarily the content of the record being retrieved (i.e., $D_{\tN}$).

The  \emph{retrieval cost} $\downloadcost (\Lambda, \storage)$ of  a  retrieval scheme $\Lambda$ (with respect to an underlying storage code $\storage$) is defined by
\[
\downloadcost (\Lambda, \storage) \triangleq \max_{k \in \K} \frac{\log |{\cal A}_{k}|}{\log |{\cal D}|}.
\]
Since each $D_{n}$ is uniformly distributed over ${\cal D}$, 
\[
\downloadcost (\Lambda, \storage) = \max_{k \in \K} \frac{\log |{\cal A}_{k}|}{H(D_{1})} \ge \max_{k \in \K} \frac{ H(A_{k})}{H(D_{1})}.
\]
The retrieval cost $\downloadcost (\Lambda, \storage)$ measures the number of bits required to retrieve from a server node, for each information bit of a record.

\begin{remark}
In this paper, we consider the case when the size of each data record  is large.
Due to \eqref{remark2.a} and \eqref{def3.eq}, $H(A_{\cal K}) \geq H(D_M)$. Thus,  the retrieval cost must grow with the size of data.
However, this is not necessary the case for upload cost (transmission bandwidths required to send the queries to the storage nodes). To illustrate, suppose we are given a $(N,K)$ storage code $\storage$, together with a retrieval scheme $\Lambda$, such that each data record is of size $\log |\cal D|$ bits. Now, suppose the size of each data record is in fact $c \log |\cal D|$ bits. In this case, we can partition each data record into $c$ chunks, each of size $\log |\cal D|$ bits. For $i=1, \ldots, c$, we may store the $i^{th}$ chunk of all the data records using the storage code $\Omega$.  To retrieve a data record, a user  can  retrieve all the $c$ chunks of the record using the retrieval scheme $\Lambda$. The catch however is that the retrieval query sent to a storage node  can be the same for every chunk. Hence, only one (not $c$) query is sent, and the upload cost stays the same for all $c$. 
In addition, similar to the case of storage cost, we use $\log |{\cal A}_{k}|$ instead of $H(A_{\cal K})$. Yet, results obtained in this paper will still hold if 
$H(A_{\cal K})$ is used.

%
%
%
\end{remark}

\subsection{Achievability}

A core question in private information retrieval is to determine  the fundamental tradeoffs between storage and retrieval costs of error-free and private retrieval schemes. To facilitate our discussion, we will need the following definitions.

\begin{df}[Achievability]
With respect to a given $K$ and collusion patterns $\tamper$, a \emph{storage-retrieval cost tuple}
$
(\storagecost^{*} , \downloadcost^{*})
$
is called $N$-achievable if there exists a $(N,K)$ storage code $\storage$  and a $\tamper$-private error free retrieval scheme $\pir$ such that 
\begin{align}
\storagecost(\storage)  & \le \storagecost^{*}  \\
\downloadcost(\Lambda, \storage)   & \le \downloadcost^{*}.  
\end{align}

Furthermore, it is called \emph{strongly achievable} if
there exists a sequence (indexed by $a$)  of  $N^{(a)}$-achievable tuples $(\storagecost^{(a)} , \downloadcost^{(a)})$ such that
\begin{enumerate}
\item $\lim_{a \to \infty}N^{(a)} = \infty$

\item
$\limsup_{a \to \infty} \storagecost^{(a)} \le \storagecost^*$

\item
$\limsup_{a \to \infty} \downloadcost^{(a)} \le \downloadcost^*$.
\end{enumerate}

\end{df}

\begin{remark} 
Clearly, a strongly achievable storage-retrieval cost tuple is also $N$-achievable (for any fixed $N$). However, as we shall prove in Theorem \ref{thm:unachievable}, the converse is not true in general.  
\end{remark}


\begin{example}\label{eg:1}
Let $D_{1}, \ldots, D_{N}$ be the data records where $D_n \in GF(q)$ for $1 \leq n \leq N$.
Construct a PIR scheme as follows:
Let $Q_{0}$ be the vector whose $\tN^{th}$ entry is one and all the others are equal to zero.
Encode the ``secret message'' $Q_{0}$  with a linear secret sharing scheme that is  defined by  $K$ random shares   $Q_{1} , \ldots, Q_{K}$ such that
\begin{enumerate}
\item
for any collusion pattern $\alpha \in \tamper$,
$
I(Q_{0} ; Q_{k} , k\in\alpha) = 0
$

\item
there exists coefficients $c_{1} , \ldots , c_{K}$ such that
\begin{align}\label{eq4}
Q_{0} = \sum_{k=1}^{K} c_{k} Q_{k}.
\end{align}
\end{enumerate}
Then,  $Q_{k}$  will be sent to storage node $k$ for $k \in \K$.
Upon receiving $Q_{k}$, node $k$ will return the response
$
A_{k} \triangleq [D_{1} , \ldots, D_{N}] Q_{k}
$.
By \eqref{eq4}, we have
\begin{align}
D_{\tN} & = [D_{1} , \ldots, D_{N}] Q_{0} \\
& =[D_{1} , \ldots, D_{N}]  \sum_{k=1}^{K} c_{k} Q_{k} \\
& =  \sum_{k=1}^{K} c_{k} A_{k}. \label{eq7}
\end{align}
Therefore, the user can decode the desired data record $D_{\tN}$. 
Furthermore, by the definition of the secret sharing code,  $I(\tN ; Q_{j}  )  = 0 $
for all $\alpha\in\tamper$. Hence, the so-constructed scheme is indeed a $\tamper$-private error-free retrieval scheme.
\end{example}

\begin{floatEq}
\begin{align}\label{eq:13}
\mV =
\left[
\begin{array}{ccc | ccc|cc|ccc}
v^{*}_{1,1,1} &  \cdots & v^{*}_{1,R,1} & v^{*}_{1,1,2} &  \cdots & v^{*}_{1,R,2} & \cdots &\cdots & v^{*}_{1,1,K} &\cdots &  v^{*}_{1,R,K} \\
\vdots  & \ddots  & \vdots & \vdots  & \ddots   & \vdots & \ddots  & \ddots & \vdots  & \ddots  & \vdots  \\
v^{*}_{L,1,1} &  \cdots & v^{*}_{L,R,1} & v^{*}_{L,1,2} &  \cdots & v^{*}_{L,R,2} & \cdots  & \cdots & v^{*}_{L,1,K} & \cdots &  v^{*}_{L,R,K} \\
v_{1,1,1} &  \cdots & v_{1,R,1} & v_{1,1,2} &  \cdots & v_{1,R,2} & \cdots & \cdots & v_{1,1,K} &  \cdots & v_{1,R,K} \\
\vdots  & \ddots  & \vdots & \vdots  & \ddots   & \vdots & \ddots  & \ddots & \vdots  & \ddots  & \vdots\\
v_{T,1,1} &  \cdots & v_{T,R,1} & v_{T,1,2} &  \cdots & v_{T,R,2} & \cdots &  \cdots & v_{T,1,K} &  \cdots & v_{T,R,K}
\end{array}
\right]
\end{align}
\end{floatEq}

The above example requires that each storage node stores every data record. Therefore, the storage cost is the highest (equal to 1). The following example however considers coded storage.

\begin{example}[PIR for coded storage]\label{eq:2}
We consider $N=2$. Each of the two record contains two bits. Assume without loss of generality that the first record is $(a_{1}, b_{1})$ and the second one is $(a_{2}, b_{2})$. There are $K=3$ storage nodes, each of which will store two bits. The first server will store $X_{1} = (a_{1}, b_{1})$, the second  $X_{2} = (a_{2}, b_{2})$ and
the third $X_{3} = (a_{1}\oplus a_{2}, a_{2} \oplus b_{2})$.
Clearly, the storage cost of this code is only $1/2$.

\begin{table}[ht]
\centering 
\begin{tabular}{|c| c| c| c|} 
\hline 
   & server {1} &  server {2} & server {3} \\ [1ex] 
\hline
\hline 
$Q_{k} = 1$ & $a_{1}$ & $a_{2}$ & $a_{1} \oplus a_{2}$\\ \hline 
$Q_{k} = 2$ & $b_{1}$ & $b_{2}$ & $b_{1} \oplus b_{2}$\\ \hline 
$Q_{k} = 3$ & $a_{1} \oplus b_{1}$ & $a_{2} \oplus b_{2}$ & $a_{1} \oplus a_{2} \oplus b_{1} \oplus b_{2}$\\ \hline 
\end{tabular}\label{table:table} 

\vspace{0.2cm}
\caption{Response mappings} 

\end{table}

Let  $\tamper$ be the set of all $\alpha$ such that $|\alpha|=1$.
We will assume that each query $Q_{k}$ (for $k=1,2,3$) is a ternary random variables taking values from the set $\{ 1,2,3\}$.
Table  I  defines how $A_{k}$ is generated from $Q_{k}$ and $X_{k}$. For example, if $Q_{2}= 2$, then $A_{2}$ (the response generated by server 2) will be equal to $b_{2}$. If $Q_{3}=3$, then $A_{3} = a_{1} \oplus a_{2} \oplus b_{1} \oplus b_{2}$.
Now, we will define the conditional probability distribution of $(Q_{1},Q_{2},Q_{3})$ given $\tN$ (the index of the record to be retrieved)
such that
$$
\Pr (Q_{1} = q_{1} , Q_{2} = q_{2}, Q_{3} = q_{3} | \tN = 1)  = 1/3
$$
if
$(q_{1},q_{2}, q_{3}) \in \{ (1,3,3) , (2,1,1) , (3,2,2)   \}$, and $0$ otherwise.
Similarly, we define
$$
\Pr (Q_{1} = q_{1} , Q_{2} = q_{2}, Q_{3} = q_{3} | \tN = 2)  = 1/3
$$
if
$(q_{1},q_{2}, q_{3}) \in \{ (3,1,3) , (1,2,1) , (2,3,2)   \}$.
It can be verified directly that
\[
I(Q_{k} ; M) = 0
\]
for all $k=1,2,3$ and hence the retrieval scheme is private.
In addition, the user can always reconstruct the desired record. For example, when $\tN=1$ and
$$
(q_{1},q_{2}, q_{3}) = (1,3,3),
$$
then the user will receive $a_{1}$, $a_{2} \oplus b_{2}$ and $a_{1} \oplus b_{1} \oplus a_{2} \oplus b_{2}$ from the three servers.
Clearly, in this case, the user can decode $a_{1}$ and $b_{1}$ back. 
The retrieval cost of this scheme is equal to $1/2$. Hence, $(1/2,1/2)$ is 2-achievable.
\end{example}

\begin{thm}\label{thm:unachievable} 
Suppose $K=3$ and 
$$\tamper = \{ \{ k \} :\: k \in \K \}.$$
Then the storage-retrieval cost tuple $(1/2,1/2)$  is 2-achievable but not strongly achievable.
\end{thm}
\begin{proof}
The 2-achievability of the storage-retrieval cost tuple $(1/2,1/2)$ follows from Example \ref{eq:2}. The proof for that it is not strongly achievable will be given in  Appendix \ref{sec:thm:unachievable}.
\end{proof}

As a corollary from Theorem \ref{thm:unachievable}, $N$-achievability and strong achievability are in fact different. This implies that the tradeoff between the storage cost and retrieval cost does depend on the number of data records in the system. To our best knowledge, this is the first proof for such a phenomenon.

\section{Linear PIR codes}\label{sec:3}

Example \ref{eg:1} is a PIR scheme constructed from  a secret sharing scheme. It requires that each storage node contains all the $N$ records and hence its storage cost is the highest (equal to 1). In contrast,   the storage cost of the code in Example 2 is only 1/2. In this section, we will propose a class of storage codes, together with its corresponding  retrieval scheme. As we shall see in Section \ref{sec:tradeoff}, our   class of codes can be constructed over a wide range of storage-retrieval cost tuples.

We will first begin with the description of the storage code. In our proposed scheme, we denote the  $N$  data records  by $\vd_{1} , \ldots, \vd_{N}$ each of which is 
a vector of length $(K-S)L$ over $GF(q)$ for some positive integer constants $L$ and $S$ (whose physical meaning will become clear later). In our proposed scheme, each data record will be ``encoded and stored'' separately. Specifically, there exists $K$ mappings 
\[
g_{k} : {\cal D} \rightarrow GF(q)^{L}, \quad k\in\K
\]
such that $g_{k}(\vd_{n})$ is the ``coded data piece'' of $\vd_{n}$ being stored in storage node $k$. Therefore, for data records   $\vd_{1} , \ldots, \vd_{N}$, 
the storage node $k$ will store the length $LN$ vector
\[
X_{k} = [g_{k}(\vd_{1}), \ldots,  g_{k}(\vd_{N})]^{\top}
\]

In order to retrieve $\vd_{n}$ without errors, it is necessary that the data record $\vd_{n}$ and 
$
[g_{k}(\vd_{n})^{\top}, k \in \K]
$
are one-to-one corresponded.  Therefore, $\vd_{n}$ and $
[g_{k}(\vd_{n})^{\top}, k \in \K]
$
can be used interchangeably. In fact, for notation simplicity, we will simply write 
$\vd_{n}$ as 
\begin{align}\label{eq:17a}
 [g_{k}(\vd_{n})^{\top}, k \in \K] 
 & = \left[
\begin{array}{ccc}
d_{n,1,1} & \cdots & d_{n,1,K} \\
\vdots & \ddots & \vdots \\
d_{n,L,1} & \cdots & d_{n,L,K}
\end{array}
\right]
\end{align}
such that 
\[
g_{k}(\vd_{n}) = [d_{n,1,k} , \ldots,  d_{n,L,k}]
\]
is the transpose of the $k^{th}$ column of $\vd_{n}$.

Now, to define the storage code, it suffices to define $g_{k}$, or equivalently,  the 
condition that $[g_{k}(\vd_{n})^{\top}, k \in \K]$   must satisfy.  
In this paper, we will consider a wide class of linear storage codes such that 
\begin{align}\label{eq:11}
\vd_{n} \mP = {\bf 0}
\end{align}
for a full-rank $K\times S$ parity check matrix
\begin{align}
\mP=
\left[
\begin{array}{ccc}
p_{1,1} &  \cdots & p_{1,S}\\
\vdots & \ddots  & \vdots \\
p_{K,1}  & \cdots  & p_{K,S}
\end{array}
\right].
\end{align}
The choice of $\mP$ determines the type of storage codes that are being used. For uncoded storage, 
$\mP$  is the $K \times K-1$ full rank matrix such that 
\[
[1, 1,\ldots, 1 ] \mP = {\bf 0}.
\]

%
%
Following our convention, the storage node  $k$ will store a length $LN$ column vector 
\[
X_{k} \triangleq [ d_{1,1,k} , \ldots, d_{1,L,k} , d_{2,1,k} , \ldots, d_{2,L,k} , \ldots, d_{N,L,k} ]^{\top},
\]
which is obtained by  concatenating the $k^{th}$ column of each record.
As a result, it is obvious that 
\begin{align}\label{eq:18a}
\sum_{k = 1}^{K} p_{k,s} X_{k}  = {\bf 0}.
\end{align}

For notation simplicity, we will refer our  code  as  $\Delta_{\mP,N,L}$ (or just $\Delta_{\mP}$ when $N$ and $L$ are  understood implicitly).
It can be directly verified that its  storage cost  is
\begin{align}\label{eq:19a}
\storagecost (\Delta_{\mP})  =  {1}/{(K-S)}.
\end{align}

Next, we will define a general class of information retrieval schemes for the above storage code $\Delta_{\mP}$. Our proposed retrieval scheme is defined by  a $(T+L) \times RK$ matrix $\mV$    (see \eqref{eq:13}), whose  columns will be respectively labeled by
\[
V_{1,1} , \ldots, V_{R,1} , V_{1,2} , \ldots, V_{R,2}, \ldots, V_{1,K} , \ldots, V_{R,K}.
\]
We will refer the retrieval scheme as $\Theta_{\mV}$.
A user can retrieve the $\tN$th record $\vd_{\tN}$ by following the  steps below: 
\begin{enumerate}
\item (Initialisation)
Generate a $LN \times T$ random matrix $ \mU$, whose entries are all randomly and independently selected from  $GF(q)$. Let the columns of  $ \mU$ be respectively 
$$U_{t}, t \in \T \triangleq \{1 \ldots, U_{T}\}.$$

\item 
(Query Generation) 
For each $k \in \K$, the user will generate $R$ ``query'' vectors $Q_{1,k} \ldots Q_{R,k}$ such that
\begin{align}
Q_{r,k} \triangleq
\sum_{t=1}^{T} v_{t,r,k} U_{t} + \sum_{\ell=1}^{L} v^{*}_{\ell,r,k}\base{\tN,\ell}.
\end{align}
Here, $v_{t,r,k}$ and $v^{*}_{\ell,r,k}$ are defined as in \eqref{eq:13} and 
$\base{\tN,\ell}$ is a length $LN$ column vector whose entry is 1 on the $L(\tN - 1) + \ell \;$th  row and is 0 otherwise.
In other words, $\base{\tN,\ell}$ is the vector such that
\begin{align}
\base{\tN,\ell}^{\top} X_{k} = d_{\tN,\ell,k}
\end{align}
for all $\tN\in\N, \ell \in \L \triangleq \{1, \ldots, L \} $ and $k\in\K$.

\item
(Response Mappings)
The collection of the $R$ query vectors 
\[
(Q_{r,k},  r \in \R \triangleq  \{1, \ldots , R\})
\]  
will then be sent  to storage node $k$, which  will respond   by computing and returning to user the tuple
$
(A_{r,k} , r=1, \ldots , R)
$
where
$
A_{r,k} =   Q_{r,k}^{\top}  X_{k} . 
$
Thus,
\begin{align}  \label{eq:17}
A_{r,k} & =  \sum_{t = 1}^{T}v_{ t, r, k}  U_{t}^{\top} X_{k} +  \sum_{\ell=1}^{L} v^{*}_{r,\ell,k} d_{\tN,\ell,k} .
\end{align}
As each storage node will return $R$ symbols back to the user,  the retrieval cost is  thus 
\begin{align}\label{eq:23a}
\downloadcost(\Theta_{\mV},\Delta_{\mP}) = \frac{R}{L(K-S)}.
\end{align}

\end{enumerate}

Ideally, the data record   $$\vd_{\tN} = (d_{\tN,\ell,k},  \ell \in \L ,  k\in\K)$$ should be obtained by solving the system of linear equations \eqref{eq:17}, together with 
\begin{align}
U_{t}^{\top} \sum_{k = 1}^{K} p_{k,s } X_{k} & = 0,  \\
\sum_{k = 1}^{K} p_{k,s }   d_{\tN,\ell,k}  & = 0 ,
\end{align}
for all $s\in {\cal S} \triangleq \{1, \ldots, S\}$ and $\ell \in \L$.
However, a retrieval scheme $\Theta_{\mV}$ is not necessarily error-free (where $\vd_{\tN}$ can be solved without errors) or private.  The following theorem gives conditions on $\mV$ (with respect to the storage code $\Delta_{\mP}$) under which our retrieval scheme is in fact error-free and private.

\begin{thm}\label{thm1}
The retrieval scheme $\Theta_{\mV}$ is error-free if the following system of linear equations
\begin{align}\label{eq:system}
\left\{
\begin{array}{l}
\sum_{k=1}^{K} w_{t,k} p_{k,s}  = 0, \hspace{2cm}  \forall t \in \T , s\in\S   \\
\sum_{k=1}^{K} w^{*}_{\ell,k} p_{k,s}  = 0, \hspace{2cm} \forall \ell \in \L, s\in\S  \\
\sum_{t = 1}^{T}v_{ t, r, k}  w_{t,k} +  \sum_{\ell=1}^{L} v^{*}_{\ell,r,k} w^{*}_{\ell,k}   =0, \\
\hspace{5cm}  \forall k\in\K,  r\in\R
\end{array}
\right.
\end{align}
has a unique solution, where the unknowns are
$$
(w_{t,k} , t \in\T, k\in\K , w^{*}_{l,k}, l\in\L , k\in\K).
$$
The condition is called \emph{retrievability condition}.

Furthermore, the retrieval scheme is  $\tamper$-private  if 
\begin{align}\label{eq:privacy condition}
\langle  V_{r, k} , r\in\R  , k\in\alpha\rangle \cap V_{0}   = \{{ \bf 0 }\}
\end{align}
for any collusion pattern $\alpha \in \tamper$.
Here,
$V_{0}$ is the $L$ dimensional vector space containing all vectors of the form
  $[ c_{1} , \ldots, c_{L} , 0 , \ldots, 0]^{\top} $. 
We call \eqref{eq:privacy condition} the   \emph{privacy condition}.
\end{thm}
\begin{proof}
For any $t \in \T$, $ r\in\R$ and  $k \in \K$, let
\begin{align}
w_{t,k} & \triangleq  U_{t}^{\top} X_{k}  \\
w^{*}_{\ell,k} & \triangleq d_{\tN,\ell,k}.
\end{align}
Then rewriting \eqref{eq:17}, we have
\begin{align}\label{thm1:eqa}
A_{r,k}  =  \sum_{t = 1}^{T}v_{ t, r, k}  w_{t,k} +  \sum_{\ell=1}^{L} v^{*}_{r,\ell,k} w^{*}_{\ell,k}.
\end{align}
Invoking \eqref{eq:18a},  for all $t\in\T$ and $s\in\S$, we have 
\begin{align}
U_{t}^{\top} \left( \sum_{k = 1}^{K} p_{k,s} X_{k}\right) = 0.
\end{align}
Consequently,
\begin{align}
\sum_{k = 1}^{K} p_{k,s }    w_{t,k} & = 0 .\label{thm1:eqb}
\end{align}
Similarly, by \eqref{eq:11}, we have
\begin{align}
\sum_{k = 1}^{K} p_{k,s }    w^{*}_{\ell,k} & = 0 , \quad \forall \ell\in\L, s\in{\cal S}. \label{thm1:eqc}
\end{align}

As the matrices $\mP$ and $\mV$ are chosen to satisfy the retrievability condition (that \eqref{eq:system} has a unique solution), the  system of linear equations \eqref{thm1:eqa}, \eqref{thm1:eqb} and \eqref{thm1:eqc} also has a unique solution. Hence,  the user can uniquely  decode $\vd_{\tN}$.  Therefore, $D_{\tN}$ is a function of $M, Q_{\K}, A_{\K}$ and hence  $\Theta_{\mV}$ is error-free.

\begin{floatEq}
\begin{align}\label{eq:25}
{\bf G} =
\left[
\begin{array}{ccc | ccc|cc|ccc}
v^{*}_{1,1,1} &  \cdots & v^{*}_{1,R,1} & v^{*}_{1,1,2} &  \cdots & v^{*}_{1,R,2} & \cdots &\cdots & v^{*}_{1,1,|\alpha|} &\cdots &  v^{*}_{1,R,|\alpha|} \\
\vdots  & \ddots  & \vdots & \vdots  & \ddots   & \vdots & \ddots  & \ddots & \vdots  & \ddots  & \vdots  \\
v^{*}_{L,1,1} &  \cdots & v^{*}_{L,R,1} & v^{*}_{L,1,2} &  \cdots & v^{*}_{L,R,2} & \cdots  & \cdots & v^{*}_{L,1,|\alpha|} & \cdots &  v^{*}_{L,R,|\alpha|} \\
v_{1,1,1} &  \cdots & v_{1,R,1} & v_{1,1,2} &  \cdots & v_{1,R,2} & \cdots & \cdots & v_{1,1,|\alpha|} &  \cdots & v_{1,R,|\alpha|} \\
\vdots  & \ddots  & \vdots & \vdots  & \ddots   & \vdots & \ddots  & \ddots & \vdots  & \ddots  & \vdots\\
v_{T,1,1} &  \cdots & v_{T,R,1} & v_{T,1,2} &  \cdots & v_{T,R,2} & \cdots &  \cdots & v_{T,1,|\alpha|} &  \cdots & v_{T,R,|\alpha|}
\end{array}
\right]
\end{align}
\end{floatEq}
\begin{floatEq}
\begin{align}\label{eq:26}
{\bf G}^{-} =
\left[
\begin{array}{ccc | ccc|cc|ccc}
v_{1,1,1} &  \cdots & v_{1,R,1} & v_{1,1,2} &  \cdots & v_{1,R,2} & \cdots & \cdots & v_{1,1,|\alpha|} &  \cdots & v_{1,R,|\alpha|} \\
\vdots  & \ddots  & \vdots & \vdots  & \ddots   & \vdots & \ddots  & \ddots & \vdots  & \ddots  & \vdots\\
v_{T,1,1} &  \cdots & v_{T,R,1} & v_{T,1,2} &  \cdots & v_{T,R,2} & \cdots &  \cdots & v_{T,1,|\alpha|} &  \cdots & v_{T,R,|\alpha|}
\end{array}
\right]
\end{align}
\end{floatEq}

Next, we will prove that $\Theta_{\mV}$ is $\tamper$-private if \eqref{eq:privacy condition} is satisfied. Let $\alpha \in \tamper$ be a collusion pattern.
Assume without loss of generality that
$\alpha = \{1, \ldots, |\alpha|\}$.
Let ${\bf G}$ and $ {\bf G}^{-}$ be two submatrices  of $\mV$  as defined
in \eqref{eq:25} and \eqref{eq:26}. Note that both ${\bf G}$ and $ {\bf G}^{-}$ have exactly $|\alpha|R$ columns.  By invoking elementary column operations, we can prove the existence of  an invertible matrix ${\bf C}$ such that ${\bf G}^{-}{\bf C}$ has rank $\delta \le T$ and
\[
{\bf G}^{-}{\bf C} =
\left[
\begin{array}{ccccccc}
f_{1,1} & f_{1,2}  &\cdots & f_{1,\delta}  & 0 & \cdots & 0 \\
\vdots &  \vdots & \ddots  & \vdots   & \vdots & \ddots & \vdots\\
f_{T,1}  & f_{T,2} & \cdots  & f_{T,\delta}  &0  & \cdots & 0
\end{array}
\right]
\]
 is in the reduced column echelon form. In particular,   for each $a=1, \ldots, \delta$, there exists an index
$1 \le t_{a} \le T$ such that
\begin{align}
f_{t_{a}, a}    & \neq  0 \label{eq:27} \\
f_{t_{a}, b}    & = 0, \quad \forall b \neq a.\label{eq:28}
\end{align}
Now, consider  the matrix
\[
{\bf G}{\bf C} =
\left[
\begin{array}{ccccccc}
f^{*}_{1,1} & f^{*}_{1,2}  &\cdots & f^{*}_{1,\delta}  & f^{*}_{1,\delta+1} & \cdots & f^{*}_{1,R|\alpha|} \\
\vdots  & \vdots   &\ddots & \vdots   & \vdots  & \cdots & \vdots  \\
f^{*}_{L,1} & f^{*}_{L,2}  &\cdots & f^{*}_{L,\delta}  & f^{*}_{L,\delta+1} & \cdots & f^{*}_{L,R|\alpha|} \\
f_{1,1} & f_{1,2}  &\cdots & f_{1,\delta}  & 0 & \cdots & 0 \\
\vdots &  \vdots & \ddots  & \vdots   & \vdots & \ddots & \vdots\\
f_{T,1}  & f_{T,2} & \cdots  & f_{T,\delta}  &0  & \cdots & 0
\end{array}
\right].
\]
The privacy condition \eqref{eq:privacy condition} guarantees that
$$
f^{*}_{\ell,\delta+1} = \cdots = f^{*}_{\ell,R|\alpha|} = 0
$$
for all $\ell = 1, \ldots, L$.
Now, let
\begin{align}\label{eq:29}
\left[Q^{*}_{1} , \ldots,  Q^{*}_{R|\alpha|}\right] & = \left[Q_{1,1} , \ldots,  Q_{R,|\alpha|}\right] {\bf C} 
\end{align}
and
$
\mU^{*} = \left[E_{\tN,1} , \cdots , E_{\tN,L}, \mU \right]
$.
Then  it can be verified directly that
\[
\left[Q_{1,1} , \ldots,  Q_{R,|\alpha|}\right]  = \mU^{*} {\bf G}.
\]
Hence,
\begin{align*}
& \hspace{-0.3cm}[Q^{*}_{1} , \ldots,  Q^{*}_{R|\alpha|}] \\
 & =[Q_{1,1} , \ldots,  Q_{R,|\alpha|}]  {\bf C} \\
& = \mU^{*} {\bf G}{\bf C}  \\
& = \mU^{*}
\left[
\begin{array}{ccccccc}
f^{*}_{1,1} & f^{*}_{1,2}  &\cdots & f^{*}_{1,\delta}  & 0 & \cdots & 0 \\
\vdots  & \vdots   &\ddots & \vdots   & \vdots  & \cdots & \vdots  \\
f^{*}_{L,1} & f^{*}_{L,2}  &\cdots & f^{*}_{L,\delta}  &0 & \cdots & 0\\
f_{1,1} & f_{1,2}  &\cdots & f_{1,\delta}  & 0 & \cdots & 0 \\
\vdots &  \vdots & \ddots  & \vdots   & \vdots & \ddots & \vdots\\
f_{T,1}  & f_{T,2} & \cdots  & f_{T,\delta}  &0  & \cdots & 0
\end{array}
\right].
\end{align*}

Note that
$Q^{*}_{i} = 0$ for all $i=\delta+1, \ldots, R|\alpha|$.
Now, for any $a=1, \ldots, \delta$,
\begin{align}
Q^{*}_{a} & = \sum_{t=1}^{T} f_{t,a} U_{t} + \sum_{\ell = 1}^{L} f^{*}_{\ell,a}\base{\tN,\ell} \\
& = f_{t_{a},a} U_{t_{a}} +  \sum_{t \neq t_{a}}  f_{t,k} U_{t} + \sum_{\ell = 1}^{L} f^{*}_{\ell,a}\base{\tN,\ell}.
\end{align}
Clearly, each $Q^{*}_{a}$ can be written as a sum of columns in $\mU^{*}$. However, by \eqref{eq:27}-\eqref{eq:28}, the column $U_{t_{a}}$ appears only in the sum for $Q^{*}_{a}$. As  entries in each $U_{t_{a}}$ are  uniformly and independently distributed $GF(q)$, we can in fact prove that
each $Q^{*}_{a}$ in $(Q^{*}_{1} , \ldots, Q^{*}_{R|\alpha|} )$
is uniformly and independently distributed to each other, no matter what   $\tN$ is. In other words, 
\begin{align}
I(Q^{*}_{1} , \ldots,  Q^{*}_{R|\alpha|} ; M) = 0.
\end{align}
Finally, by \eqref{eq:29}, we have 
\begin{align}\label{eq:40}
I(Q_{1,1} , \ldots,  Q_{R,|\alpha|} ; M) = 0.
\end{align}
As \eqref{eq:40} holds for all $\alpha \in \tamper$,  our retrieval scheme is indeed $\tamper$-private. And the theorem is proved. 
\end{proof}

\begin{remark}
It is worth to mention that our codes and retrieval schemes are well-defined irrespective of $N$. Therefore, if a storage-retrieval cost tuple 
$
(\storagecost , \downloadcost )
$
can be achieved using our proposed schemes, the tuple is also strongly achievable. 
\end{remark}

\section{Tradeoff analysis}\label{sec:tradeoff}

In the previous section, we proposed a linear storage code $\Delta_{\mP}$ and retrieval scheme $\Theta_{\mV}$. Conditions under which the retrieval scheme is error-free and private have also been identified in Theorem \ref{thm1}.  In this section, we will analyse the tradeoffs between the storage cost and the retrieval cost of our codes in  the simplest special case when 
$$\tamper = \{ \{ k \} :\: k \in \K \}.$$

\begin{prop}\label{prop:converse}
If   $\mV$ and $\mP$ satisfy the retrievability condition in Theorem \ref{thm1}, then  for any subset $\beta \subseteq \K$,
$$
(T+L-R)(K-|\beta|) \le \rank{\mP(\beta)}(T+L)
$$
where
$\mP(\beta)$ is the  submatrix of $\mP$ by removing the rows indexed by
$k \in \beta$. 
\end{prop}
\begin{proof}
Let $\beta$ be a subset of $\cal K$. Consider again the system of equations in \eqref{eq:system}. 
Suppose the user is informed by a genie about the values of the following unknowns
\begin{align}\label{eq:prop1a}
\{(w_{t,k}, w^{*}_{\ell,k}) :   \: t\in \T, \ell \in \L \text{ and } k \in \beta  \}.
\end{align}
In that case,  the number of remaining unknowns is now only 
$
(T+L) (K - |\beta|)
$.
Given \eqref{eq:prop1a}, the following equations  
\begin{align}
\sum_{t = 1}^{T}v_{ t, r, k}  w_{t,k} +  \sum_{\ell=1}^{L} v^{*}_{\ell,r,k} w^{*}_{\ell,k}  & =0, \quad \forall k\in\beta,  r \in \R
\end{align}
will become redundant.  By direct counting,  there are at most
\[
\rank{\mP(\beta)}(T+L) + (K - |\beta|)R
\]
non-redundant equations left in \eqref{eq:system}.
Therefore, if the linear system \eqref{eq:system}  has a unique solution, then
\begin{align}
(T+L) (K - |\beta|) \le \rank{\mP(\beta)}(T+L) + (K - |\beta|)R,
\end{align}
or equivalently,
\begin{align}
(T+L-R)(K-|\beta|) \le \rank{\mP(\beta)}(T+L).
\end{align}
The proposition is thus proved.
\end{proof}

\begin{prop}\label{prop:converse2}
If the matrix $\mV$  satisfies the privacy condition \eqref{eq:privacy condition}, then $R \le T$.
\end{prop}
\begin{proof}
Consider any  $k \in \K$ and  the columns
\begin{align}
\left[
\begin{array}{c}
v^{*}_{1,1,k}  \\
\vdots  \\
v^{*}_{L,1,k} \\
v_{1,1,k} \\
\vdots  \\
v_{T,1,k}
\end{array}
\right]
\cdots \cdots
\left[
\begin{array}{c}
v^{*}_{1,R,k}  \\
\vdots  \\
v^{*}_{L,R,k} \\
v_{1,R,k} \\
\vdots  \\
v_{T,R,k}
\end{array}
\right]
\end{align}
in $\mV$.
We may  assume without loss of generality that all the columns are independent (otherwise, some of the columns can be discarded).
Now,   if $R > T$, then
\[
\rank{\langle  V_{r, k} , r\in\R  , k\in\alpha\rangle} > T.
\]
As  
$
\rank{ \langle  V_{r, k} , r\in\R  , k\in\alpha , V_{0} \rangle  }  \le T+L
$
and 
$\rank{V_{0}} = L$,
we have  $
\rank{ \langle  V_{r, k} , r\in\R  , k\in\alpha\rangle \cap V_{0} }  > 0. 
$
The proposition is thus proved.
\end{proof}

\begin{thm}[Tradeoff]\label{thm:tradeoff}
For any storage code $\Delta_{\mP}$ and retrieval scheme $\Theta_{\mV}$   satisfying the retrievability and privacy condition,
\begin{align}\label{eq:thm3}
1  \le  \downloadcost (\Theta_{\mV}, \Delta_{\mP}) \left(K - \frac{1}{\storagecost (\Delta_{\mP})} \right).
\end{align}
Equivalently,
\begin{align}
\downloadcost (\Theta_{\mV}, \Delta_{\mP})   \ge \frac{\storagecost (\Delta_{\mP})}{ K \storagecost (\Delta_{\mP}) - 1}. 
\end{align}

\end{thm}
\begin{proof}
Let $\beta = \emptyset$. Applying Proposition \ref{prop:converse}, we have
\begin{align}
(T+L-R)K & \le S(T+L) \\
T(K-S) & \le RK - L(K-S) \\
T & \le \frac{ RK }{K-S}- L .
\end{align}
On the other hand, from Proposition \ref{prop:converse2}, we have
$R \le T$. Therefore, we have
\begin{align*}
R & \le \frac{ RK }{K-S}- L,  \\
\frac{R}{L} & \le \frac{ RK }{L(K-S)}- 1 .
\end{align*}

By \eqref{eq:19a} and \eqref{eq:23a}, we have
%
\begin{align*}
1 & \le \frac{ RK }{L(K-S)}  - \frac{R}{L} \\
  & = \frac{ R  }{L(K-S)}  \left(K  - \frac{K-S}{1} \right).
 \end{align*}
 Therefore, 
\[ 
1   \le  \downloadcost (\Theta_{\mV}) \left(K - \frac{1}{\storagecost (\Delta_{\mP})} \right)
\]
\end{proof}

\begin{cor}\label{cor1}
When $K=3$, there does not exist any storage code $\Delta_{\mP}$  and retrieval scheme $\Theta_{\mV} $ satisfying the retrievability and privacy condition such that
\[
\downloadcost (\Theta_{\mV}, \Delta_{\mP})  = \storagecost (\Delta_{\mP}) = 1/2.
\]
\end{cor}
\begin{proof}
A direct consequence of Theorem \ref{thm:tradeoff}.
\end{proof}

Theorem \ref{thm:tradeoff} gives a  tradeoff between storage cost and retrieval cost (in the context of our proposed classed of schemes). It has been well known that maximum distance separable (MDS) code has the lowest storage cost (compared to other storage codes with the same reliability). In the following theorem, we will show that the optimal storage-retrieval cost tradeoff can be achieved for MDS storage codes.

\begin{thm}[Optimality]\label{thm:achievabilityMDS} 
Consider any fixed $K$, $S$. 
One can construct a $(K,K-S)$  maximum distance separable (MDS) code (specified by the parity check matrix $\mP$) and an error-free, private retrieval scheme $\Theta_{\mV}$  such that the lower bound \eqref{eq:thm3} is tight. In other words,
\[
\downloadcost (\Theta_{\mV}, \Delta_{\mP})   = \frac{\storagecost (\Delta_{\mP})}{ K \storagecost (\Delta_{\mP}) - 1}. 
\]
\end{thm}
\begin{proof}
See Appendix \ref{appendix:proof:thm:achievabilityMDS}. 
\end{proof}


\section{Conclusion}

Existing work on private information retrieval (PIR) problems  largely focused on uncoded data storage (where every storage node stores all
the data records). Their main focus was to design a retrieval scheme which has the 
lowest total upload (for transmitting retrieval queries) and download (for data retrieval) costs. 
One main drawback for uncoded storage is its high storage cost that the same data copy will be stored for multiple times. 
In this paper, we studied the problem of private information retrieval for coded data storage. Another major difference between this paper and existing works
is that we focus only on the download cost. We have justified that the upload cost does not increase with the data record size. Therefore, 
in the scenario where data record size is  large,   the upload cost can become negligible. 
A fundamental question in PIR problems in coded storage is the characterisation of the tradeoffs between storage costs and retrieval costs.
We have shown, to our surprise, that the tradeoff  in fact depends on the number of data records in the system.

In the second half of the paper, we proposed a fairly general class of linear storage codes (where the underlying storage codes are specified by a parity check matrix) and an associated information retrieval scheme. We have identified conditions under which the retrieval scheme is error-free and private. 
One interesting property of our scheme is that  the storage and retrieval costs are  the same for any number of data records. 
Tradeoffs between the storage cost and the retrieval cost of our proposed schemes are then obtained. 
We also considered specific scenarios where MDS codes are used in the underlying coded storage. We have showed that a ``randomly constructed'' retrieval scheme is in fact optimal (in minimising the retrieval costs).



\bibliographystyle{IEEEtran}
\bibliography{StorageNetwork}

\begin{appendices}

\section{Proof  of Theorem \ref{thm:unachievable}}\label{sec:thm:unachievable}

Suppose to the contrary that the storage-retrieval cost tuple $(1/2 , 1/2)$ is strongly achievable. By definition, there exists a sequence  of $(N^{(a)} , 3)$ storage codes  
$
\storage^{(a)}  
$ and 
$\tamper$-private, error-free retrieval schemes 
$
\Lambda^{(a)} = (\Lambda_{1}^{(a)}, \Lambda_{2,k}^{(a)})
$
such that 
 \begin{align}
\limsup_{a\to\infty} \storagecost (\storage^{(a)} )  & \le \frac{1}{2}  \label{appendix5.proof.a}\\
\limsup_{a \to\infty} \downloadcost (\Lambda^{(a)}, \storage^{(a)} )  & \le \frac{1}{2} \label{appendix5.proof.b}\\
\lim_{a \to \infty} N^{(a)} & = \infty. \label{appendix5.proof.c}
\end{align}
 
For each $a$, let $(D^{(a)}_{n}, n = 1, \ldots, N^{(a)})$ be  the data records and
$(X^{(a)}_{k}, k=1, \ldots, K^{(a)})$ be the content stored at the storage nodes.
The retrieval scheme is defined by $(M^{(a)}, Q^{(a)}_{k} , A^{(a)}_{k}, k\in \K )$.
For notation simplicity, we will often drop the superscript $^{(a)}$ in the remainder of the proof.

First, notice that $(D_{n}, n\in \N)$ is mutually independent (see also \eqref{remark2.d}). Therefore, for any $k\in\K$,
\begin{align}
\I{Q_{k}A_{k}}{D_{\N}}
\ge 
\sum_{n\in\N}\I{Q_{k}A_{k}}{D_{n}}.
\end{align}
On the other hand, 
\begin{align}
\I{Q_{k}A_{k}}{D_{\N}} 
& = \I{Q_{k}}{D_{\N}} + \I{A_{k}}{D_{\N} | Q_{k}}  \\
& \le H(A_{k} | Q_{k} ).
\end{align}
Here, the last inequality follows from \eqref{remark2.a}. 
Therefore
\begin{align}\label{eq78}
H(A_{k} |Q_{k}) \ge \sum_{n\in\N}\I{Q_{k}A_{k}}{D_{n}}
\end{align}
and consequently, 
\begin{align}
& \hspace{-0.5cm} \limsup_{a\to\infty} \frac{ \max_{k\in\K} \sum_{n=1}^{N} \I{Q_{k}A_{k}}{D_{n}}  }{ H(D_{1})}  \nonumber \\
& \le \limsup_{a \to\infty} \frac{ \max_{k\in\K} H(A_{k}|Q_{k})  }{ H(D_{1})} \\
 & \le \frac{1}{2} 
\end{align}
where the last inequality follows from \eqref{appendix5.proof.a}.
Hence, for any $c > 1/2$ and sufficiently large $a$, we have 
\begin{align}
c \ge \frac{ \max_{k\in\K} \sum_{n=1}^{N} \I{Q_{k}A_{k}}{D_{n}}  }{ H(D_{1})}.
\end{align}
Consequently, 
\begin{align}
 \frac{\sum_{k\in\K}\sum_{n=1}^{N}  \I{Q_{k}A_{k}}{D_{n}} }{N} 
& \le \frac{c K H(D_{1})}{N}.
\end{align}
Therefore, there must exist at least one $n^{*}$ such that 
\[
\sum_{k\in\K}\I{Q_{k}A_{k}}{D_{n^{*}}} \le \frac{c K H(D_{1})}{N}
\]
which implies that  for all $k\in\K$, 
\begin{align}
\I{Q_{k}A_{k}}{D_{n^{*}}} \le \frac{c K H(D_{1})}{N}. 
\end{align}

Now, notice that 
\begin{align*}
& \hspace{-0.5cm} H(M, Q_{k} ,X_{k},  D_{n^{*}},A_{k}) \nonumber\\
&  = H(M) + H(Q_{k}|M )+ H(X_{k} D_{n^{*}} | M,Q_{k}) \nonumber \\
& \qquad   + H(A_{k}| M,Q_{k},X_{k}, D_{n^{*}}) \\
& \nequal{(i)} H(M) + H(Q_{k})+ H(X_{k} D_{n^{*}} | Q_{k}) \nonumber \\
& \qquad   + H(A_{k}| Q_{k},X_{k}, D_{n^{*}}) \\
& = H(M) + H(Q_{k} ,X_{k},  D_{n^{*}},A_{k})
\end{align*}
where 
$(i)$ follows from \eqref{remark2.a}, \eqref{remark2.c} and \eqref{def4:eq.a}. 
Hence, $M$ and $(Q_{k} ,X_{k},  D_{n^{*}},A_{k})$ are independent. Consequently, for sufficiently large $a$, 
\begin{align}
I(D_{n^{*}} ; Q_{k} A_{k} | M=n^{*}) 
& = I(D_{n^{*}} ; Q_{k} A_{k} ) \\
& \le \frac{c K H(D_{1})}{N}  \\
& \nequal{(ii)} \frac{c K H(D_{n^{*}})}{N}.  \label{eq84a}
\end{align}
Here, $(ii)$  follows from that $H(D_{n})$ is constant for all $n$.
Similarly, by using \eqref{appendix5.proof.b} and the privacy condition \eqref{def4:eq.a}, we can prove that for  sufficiently large $a$, 
\begin{align}
H(A_{k} | Q_{k}, M=n^{*})  & \le   c H(D_{n^{*}}).  \label{eq84}
\end{align}

In addition, by  \eqref{remark2.a}, \eqref{remark2.c} and \eqref{def3.eq}, we have  
\begin{align}
I(X_{\K}, D_{n^{*}} ; Q_{\K} |M=n^{*} ) & = 0 \label{eq:75a}\\
H(D_{n^{*}} | Q_{\K } , A_{\K } , M=n^{*}) & = 0  \label{eq:76a}\\  
H(A_{k} | Q_{k},X_{k}, M=n^{*})   & =  0 \label{eq:77a}
\end{align}

Before we continue, we will need the following intermediate lemma.
\begin{lemma}\label{lemma2}
\begin{multline}
 2  H(D_{n^{*}} )  + H(A_1| Q_{\K}A_2  D_{n^{*}} ) \\
\le     2 H(A_1| Q_1 )  +   H(A_2| Q_2 X_2 )    + 2 H(A_3| Q_3 )  \\ 
+    H(D_{n^{*}}| Q_{\K} A_{\K} ) +   I(Q_2A_2; D_{n^{*}} )  
  + I(Q_{\K}; X_2 D_{n^{*}}). \label{lemma1:eqa}
\end{multline}
Consequently, 
\begin{align}
  &2  H(D_{n^{*}}| M=n^{*} )   + H(A_1| Q_{\K}A_2 D_{n^{*}}, M=n^{*}) \nonumber\\
\le&     2 H(A_1| Q_1 ,  M=n^{*})   +   H(A_2| Q_2 X_2,  M=n^{*} )   \nonumber\\
 &+ 2 H(A_3| Q_3 , M=n^{*})  +    H(D_{n^{*}}| Q_{\K} A_{\K}, M=n^{*} ) \nonumber\\ 
 &+   I(Q_2A_2; D_{n^{*}} | M=n^{*})   + I(Q_{\K}; X_2 D_{n^{*}}| M=n^{*}). \label{lemma1:eqb}
\end{align}
\end{lemma}
\begin{proof}
Using \cite{HoTanYeung14}, we can obtain the following equality 
which can be directly verified by re-writing each term in terms of joint entropy.  
\begin{align*}
&  \hspace{-0.5cm}  -2  H(D_{n^{*}} )   - H(A_1| Q_{\K} A_2 D_{n^{*}} ) \nonumber\\
=& H(A_2| Q_{\K}, D_{n^{*}}, A_1, A_3, X_2 ) +
  I(A_1; Q_3| Q_1, Q_2, A_2 ) \nonumber\\
&  + H(A_3| Q_{\K} A_1 A_2 D_{n^{*}} X_2 ) +   I(A_1; A_2| Q_{\K} D_{n^{*}} X_2 )  \nonumber\\
& + I(A_1; Q_2| Q_1 A_2 ) +  I(A_1; A_2| Q_1 ) +   I(A_1; A_3| Q_{\K} A_2 ) \nonumber\\
&  + I(Q_1; A_2| Q_2 Q_3 ) + I(Q_2; D_{n^{*}}| X_2 ) +   I(Q_2; X_2 )  \nonumber\\
& + I(Q_1; A_3| Q_3 A_2 ) +   I(A_2; A_3| Q_3 ) +   I(Q_3; A_2| Q_2 )  \nonumber\\
& + I(Q_2; A_3| Q_1 Q_3 A_2 ) + I(Q_3; X_2| Q_1 Q_2 A_2 D_{n^{*}} )  \nonumber\\
& + I(A_2; D_{n^{*}}| Q_2 X_2 ) +  I(A_2; A_3| Q_{\K} A_1 X_2 D_{n^{*}} )  \nonumber\\
& + I(Q_1; X_2| Q_2 A_2 D_{n^{*}} ) +  I(X_2; A_3| Q_{\K} A_1 A_2 D_{n^{*}} )  \nonumber\\
&   - 2 H(A_1| Q_1 )   -   H(A_2| Q_2 X_2 ) - 2 H(A_3| Q_3 )  \nonumber\\
&    -    H(D_{n^{*}}| Q_{\K}  A_{\K} )   -   I(Q_2  A_2; D_{n^{*}} )  - I(Q_{\K}; X_2 D_{n^{*}})  . 
\end{align*}
Hence, the lemma follows from the nonnegativity of mutual information and entropies.
\end{proof}

Now, by invoking Lemma \ref{lemma2}, and  \eqref{eq84a} -- \eqref{eq:77a}, we have
\begin{multline*}
H(A_1| Q_{\K}A_2 D_{n^{*}}, M = n^*)  \leq  \left(4c-2 + \frac{cK}{N}\right) H(D_{n^{*}}).
\end{multline*}
Similarly, for $i\neq j$, 
\begin{multline*}
H(A_i| Q_{\K}A_j D_{n^{*}}, M = n^*)  \leq  \left(4c-2 + \frac{cK}{N}\right) H(D_{n^{*}})
\end{multline*}
by symmetry. 
Finally, for any $n\in\N$,
\begin{align}
& \hspace{-0.5cm} H(D_{n} | X_1, D_{n^{*}}) \nonumber \\
&  \nequal{(iii)} H(D_{n} | X_1, D_{n^{*}}, Q_\K, M=n)\\
&\le H(D_{n}, A_\K  | X_1, D_{n^{*}}, Q_\K, M=n ) \\
&\nequal{(iv)} H( A_\K  | X_1, D_{n^{*}}, Q_\K,M=n ) \\ 
&\le \sum_{k \neq 1} H( A_{k} | X_1, A_{1}, D_{n^{*}}, Q_\K, M=n)  \nonumber \\
& \quad \quad + H( A_{1} | X_1,  D_{n^{*}}, Q_\K, M=n)  \\
&= \sum_{k \neq 1} H( A_{k} | X_1, A_{1}, D_{n^{*}}, Q_\K, M=n)  \nonumber \\
& \le K \left(4c-2 + \frac{cK}{N}\right) H(D_{n^{*}}) 
\end{align}
where $(iii)$ follows from \eqref{remark2.a} and $(iv)$ from \eqref{def3.eq}.
Hence, 
\begin{align}
H(D_{\N} | X_1, D_{n^{*}} ) & \le \sum_{n\in\N} H(D_{n} | X_1, D_{n^{*}}) \\
& \le NK \left(4c-2 + \frac{cK}{N}\right) H(D_{n^{*}}). 
\end{align}
Consequently, 
\begin{align*}
& \hspace{-0.4cm} N H(D_{n^{*}})   \\
& = H(D_{\N})   \\
& \le H(  X_1, D_{n^{*}} )  + H(D_{\N} | X_1, D_{n^{*}} ) \\
& \le H(  X_1) + H( D_{n^{*}} )  + H(D_{\N} | X_1, D_{n^{*}} ) \\
& \le H(  X_1) + H( D_{n^{*}} )  +  NK \left(4c-2 + \frac{cK}{N}\right) H(D_{n^{*}}).
\end{align*}
Therefore, 
\begin{align}
1 \le \frac{H(X_{1})}{N H(D_{n^{*}})} + \frac{1}{N} + K \left(4c-2 + \frac{cK}{N}\right).
\end{align} 
Consequently, 
\[
\liminf_{a\to\infty}  \frac{H(X_{1})}{N H(D_{n^{*}} )} + \frac{1}{N } + K \left(4c-2 + \frac{cK}{N}\right) \ge 1.
\]
As $c$ is arbitrarily close to $1/2$, together with \eqref{appendix5.proof.c}, we have 
\[
 \liminf_{a\to\infty}  \frac{H(X_{1})}{N H(D_{n^{*}} )}  \ge 1.
\]
In other words, $\storagecost(\storage^{(a)}) \ge 1$. The theorem is proved.

\section{Proof of Theorem \ref{thm:achievabilityMDS}}\label{appendix:proof:thm:achievabilityMDS}

Consider any $K$ and $S < K$. For sufficiently large field $GF(q)$, we can always construct a $(K, K-S)$  MDS code. Let its $K \times S$ parity check matrix be $\mP$ and $\Delta_{\mP}$ be the corresponding induced storage code. Hence, 
\[
\storagecost(\Delta_{\mP}) = \frac{1}{K-S}.
\]

Let $R=T=K-S$ and $L=S$. Hence, $T+L = K$ and the matrix $\mV$ is indeed a square matrix. Now consider the retrieval scheme $\Theta_{\mV}$. It can be verified easily that 
\[
\downloadcost (\Theta_{\mP},\Delta_{\mV}) = \frac{1}{S}
\]
and hence the equality in \eqref{eq:thm3} holds.

To prove the theorem, it suffices to prove that one can construct a matrix $\mV$ such that its induced retrieval scheme $\Theta_{\mV}$ is error-free and private. First of all, for notation simplicity, we rewrite the variables in \eqref{eq:system} as follows:
\begin{align}
z_{i,k} = 
\begin{cases}
w^{*}_{i,k} & \text{ if }  1 \le i \le L \\
w_{i-L,k} & \text{ if }  1 \le i - L \le T. 
\end{cases}\\
y_{i,r,k}=
\begin{cases}
v^{*}_{i,r,k} & \text{ if } 1 \le i \le L \\
v_{i-L,r,k} & \text{ if }  1 \le i - L \le T 
\end{cases}\label{eq54a}
\end{align}

Using the new notation,  \eqref{eq:system} becomes  
\begin{align}
\sum_{k=1}^{K}  p_{k,s} z_{i,k}  = 0, &  \quad \forall i \in \K , s\in\S  \label{appendixa:eq1} \\
\sum_{i=1}^{K} y_{i,r,k} z_{i,k}   =0, & \quad \forall k\in\K,  r\in\R. \label{appendixa:eq2}
\end{align}

As $\mP$ is the  $K \times S$ parity-check matrix of a $(K, K-S)$ MDS code. Hence, for any $i\in \K$ and a subset $\beta \subseteq \K$ of size at least  $K-S$, if 
\[
\sum_{k=1}^{K}  p_{k,s} z_{i,k}  = 0, \quad \forall s\in\S
\]
and
$z_{i,k} = 0$ for all $k \in \beta$, then $z_{i,k} = 0$ for all $k\in\K$.

Next, let  
\begin{align}\label{eq57a}
y_{i,r,k} \triangleq  
\begin{cases}
1 & \text{ if }   0 = k - i - r  \mod K   \\
0 & \text{ otherwise}
\end{cases}
\end{align}
for $i,k \in \K$ and $r\in \R$.  
Let 
\[
\Box \triangleq \{ (i,k) : \: \exists r \in \R \text{ such that } 0 = k - i - r  \mod K   \}.
\]
Then \eqref{appendixa:eq2} becomes 
\[
z_{i,k} = 0 , \quad \forall  (i,k) \in \Box.
\]
Also,  for any fixed $i \in \K$, the set 
$
\{ k: (i,k) \in\Box \}
$
has exactly $R = K-S$ elements. Therefore, by the construction of $\mP$, we can conclude that 
$
z_{i,k} = 0 
$
for all $i\in \K$. 
Hence, for our choice of $y_{i,r,k}$ (and hence, $v^{*}_{i,r,k}$ and $v_{i,r,k}$), the retrievability condition is satisfied.

So far, we  proved the existence of $\mV$ (via \eqref{eq54a} and \eqref{eq57a}) such that the retrievability condition  will be satisfied. However, the chosen $\mV$ may not satisfy the privacy condition \eqref{eq:privacy condition}. In the following, we will prove the existence of a matrix $\mV$ which can satisfy both conditions.

The proof for existence relies much on the well-known Schwartz-Zippel Lemma~\cite{schwartz1980fast,zippel1979probabilistic} which is restated as follows.

\begin{lemma}[Schwartz-Zippel Lemma~\cite{schwartz1980fast,zippel1979probabilistic}]
Let $F[b_{1}, \ldots, b_{n}]$ be a non-zero polynomial with degree $\gamma \ge 0$ over the finite field $GF(q)$. If $b_{1}, \ldots, b_{n}$ are randomly and independently chosen from the field, then the probability that $F[b_{1}, \ldots, b_{n}] =0$ will be at most $d/q$. 
\end{lemma}

Now, consider again the system of linear equations \eqref{eq:system}.  The retrievability condition will be satisfied if and only if the determinant of the coefficient matrix for \eqref{eq:system} is non-zero. Let 
\[
F(v^{*}_{\ell,r,k}, v_{t,r,k}, t\in\T, r\in\R, k\in\K, \ell\in\L)
\]
be the determinant function. Here, we assume that $\mP$ is given and fixed.
Clearly, the determinant function $F$ is a non-zero polynomial because we have already shown earlier that it has non-zero values for some choices of  
\[
v^{*}_{\ell,r,k}, v_{t,r,k}, t\in\T, r\in\R, k\in\K, \ell\in\L.
\]

%
%

Applying the Schwartz-Zippel Lemma, for sufficiently large $q$,  the determinant function is non-zero with arbitrarily high probability. In other words, the retrievability condition holds with high probability, if 
\[
v^{*}_{\ell,r,k}, v_{t,r,k}, t\in\T, r\in\R, k\in\K, \ell\in\L
\]
are  in fact chosen randomly (and when $q$ is sufficiently large). 
Finally,  to prove the theorem, it suffices to prove that the privacy condition
will also be satisfied by a randomly generated $\mV$ 
with arbitrarily high probability.

Notice  the columns 
\[
V_{r,k} = [ v^{*}_{1,r,k},  \ldots, v^{*}_{L,r,k}, v_{1,r,k}, \ldots,  v_{T,r,k} ]^{\top}.
\]
Since each element $v^{*}_{\ell,r,k}$ or $v_{t,r,k}$  is  randomly and independently selected and $\rank {V_{0}} = S$,   the probability that 
\begin{align}
\langle   V_{r,k}, r = 1, \ldots R \rangle \cap V_{0} = \{0\}
\end{align}
is equal to 
\[
\prod_{c=0}^{K-S-1} \left( 1 - \frac{q^{S+c}}{q^{K}} \right) \ge \left(1 - \frac{1}{q} \right)^{K-S}.
\]
Hence, the probability  can be made arbitrarily close to 1  for sufficiently large $q$. In other words,   the privacy condition will also be satisfied by a randomly generated $\mV$. 
The theorem is proved.

\end{appendices}

\end{document}